\documentclass[10pt,conference]{IEEEtran}

\usepackage[utf8]{inputenc}
\usepackage[T1]{fontenc}

\PassOptionsToPackage{hyphens}{url}\usepackage{hyperref}

\IEEEoverridecommandlockouts
\usepackage{cite}
\usepackage{amsmath,amssymb,amsfonts}
\usepackage{algorithmic}
\usepackage{graphicx}
\usepackage{textcomp}
\usepackage{tabularx}
\usepackage{booktabs}
\usepackage{multirow}
\usepackage{url}
\usepackage{paralist}
\usepackage{flushend}
\usepackage{tikz}
\usepackage{xcolor}
\def\BibTeX{{\rm B\kern-.05em{\sc i\kern-.025em b}\kern-.08em
    T\kern-.1667em\lower.7ex\hbox{E}\kern-.125emX}}

\usepackage{listings}
\lstdefinestyle{listingstyle}{
    frame=single,
    basicstyle=\ttfamily\footnotesize,
    breakatwhitespace=false,
    breaklines=true,
    captionpos=b,
    keepspaces=true,
    showspaces=false,
    showstringspaces=false,
    showtabs=false,
    tabsize=2
}
\newcommand\copyrighttext{%
  \footnotesize \textcopyright~2026 IEEE. Personal use of this material is permitted. Permission from IEEE must be obtained for all other uses, in any current or future media, including reprinting/republishing this material for advertising or promotional purposes, creating new collective works, for resale or redistribution to servers or lists, or reuse of any copyrighted component of this work in other works.}
\newcommand\copyrightnotice{%
\begin{tikzpicture}[remember picture,overlay]
\node[anchor=south,yshift=10pt] at (current page.south) {\fbox{\parbox{\dimexpr\textwidth-\fboxsep-\fboxrule\relax}{\copyrighttext}}};
\end{tikzpicture}%
}

\begin{document}

\title{On the Viability of Requirements Generation From Code: An Experience Report}

\author{\IEEEauthorblockN{Alexander Korn, Jone Bartel, Max Unterbusch, Andreas Vogelsang}
\IEEEauthorblockA{\textit{paluno -- The Ruhr Institute for Software Technology} \\
\textit{University of Duisburg-Essen} \\
Essen, Germany \\
\{firstname\}.\{lastname\}@uni-due.de}
}

\maketitle

\begin{abstract}
Empirical research in Requirements Engineering is hampered by a lack of adequate datasets that pair source code with corresponding requirements.
A tempting route to addressing this lack is the use of Large Language Models to synthesize requirements from existing code bases.
We investigate this question by evaluating an LLM-based and RAG-supported agentic approach that generates requirements from source code, verifies their implementation status relying on a human-in-the-loop, and synthetically introduces requirements smells and non-implemented requirements. Our goal was to create datasets that mimic reality and foster empirical RE research.
However, during the study, various problems arose, leading to this experience report.
Contrary to our initial hypotheses, LLMs were unable to (i) generate non-implemented requirements reliably, (ii) generate high quality requirements, and (iii) reliably introduce synthetic requirements smells.
Furthermore, neither an LLM nor a single human-in-the-loop suffices to detect requirements smells reliably.
These findings suggest that the generation of code-to-requirements datasets using LLMs is not yet viable and requires human supervision, especially for quality assurance.
We critically reflect on our lessons learned and draw relevant conclusions for both researchers and practitioners.
\end{abstract}

\begin{IEEEkeywords}
Agentic AI, LLMs, Synthetic Datasets, Requirements Smells
\end{IEEEkeywords}

\copyrightnotice

\section{Introduction}

Empirical research in \textit{Requirements Engineering (RE)} is based on the quality of the data used.
To best support research, data should (i) be FAIR (findable, accessible, interoperable, reusable)~\cite{Lamprecht2019}, (ii) contain a realistic number of artifacts, (iii) contain high-quality requirements, and yet (iv) be realistic in quality~\cite{yang2026assessing}, so that real-world imperfections can also be researched.
For supporting research on the relation between requirements and code~\cite{Guo2025}, datasets should also (v) contain relevant trace links, e.g. requirements-to-code traces.
Although there are datasets satisfying some of these dimensions, there is a lack of datasets satisfying all dimensions (cf. Section~\ref{sec:related-work}).
Ideally, industry datasets could be used, which would render (iv) useless.
However, industry data can rarely be disclosed for proprietary or legal reasons, leading to a large number of academic or toy datasets used in RE research~\cite{Motger26}.

Due to significant advances in the performance of \textit{Large Language Models (LLMs)}, researchers started using LLMs to generate or synthesize textual artifacts such as code and requirements~\cite{jiang2026survey, el2025synthline} or even requirements traceability data~\cite{xu2024bridging, dearstyne2025intelligent, wang2026r2code}.
However, the generated datasets show deficiencies in quality and realism.

Our initial idea was to propose an agentic human-in-the-loop approach that synthesizes requirements from existing code bases and link them to the corresponding code locations.
Specifically, we wanted to use Retrieval-Augmented Generation (RAG) to ground the requirements in code, presumably leading to high-quality requirements with clear links to code.
We operationalize requirements quality by using requirements smells as quantifiable proxies for quality issues in requirements~\cite{Femmer2017,gentili2023characterizing}.
We hypothesized that the generated requirements would show a low smell rate and are, in fact, implemented in the source code.
We further hypothesized that we could extend the approach to generate a control group of non-implemented requirements with similar properties to the implemented requirements.
Finally, we planned to extend the approach by an agent that introduces requirements smells synthetically, such that a realistic dataset would be generated with explicit labeling of requirements smells.

In this experience report, we describe how we implemented and evaluated this approach and finally had to withdraw most of our hypothesis. In contrast, we observe the following problems:
\begin{compactitem}
    \item The supposedly high-quality requirements still exhibited a smell rate of 24.5\%.
    \item 73.8\% of the smell-free but \textit{non-implemented} requirements generated were marked as \textit{implemented} by a human evaluator, showing that the LLM often generates implemented requirements instead.
    \item The inter-rater agreement for smell detection and appropriateness evaluation of generated requirements showed fair agreement at best.
    \item Smell evaluation by the human-in-the-loop was highly subjective and context dependent.
\end{compactitem}

In this experience report, we describe our approach, the evaluation, the results, and reflect on these. The key contributions are:
\begin{compactitem}
    \item \textbf{An empirical evaluation} of our approach across 2 software projects, examining implementation accuracy, hallucination rate, generation quality, and human-LLM reliability on smell detection.
    \item \textbf{Concrete lessons learned} on the limitations of LLM-based requirements generation, specifically around smell labeling subjectivity, the LLM complying with the given prompts, and codebase-dependency of the results.
    \item \textbf{Two publicly available requirements-to-code datasets} generated and peer-reviewed during the study.
\end{compactitem}

\section*{Data Availability}

All data used in this study, including the requirements-to-code datasets generated during the evaluation, the experimental implementation of the presented approach, 
and the analysis code used to process the study results, are available in the 
supplementary material\footnote{Supplementary material: \url{https://doi.org/10.6084/m9.figshare.32393787}}.

\section{Related Work}
\label{sec:related-work}

Related work stems from two areas: requirement-to-code-traceability and the more recent requirement-from-code generation.

\subsection{Requirement-to-code Datasets}

Requirement-and-code datasets provide requirements along with the specific code traces that implement them.
Although their applicability is broad, the most common use is by the traceability community interested in creating or recovering traceability links from requirements to source code~\cite{Guo2025}.
Zoogan~et~al.~\cite{zogaan2017datasets} list all datasets used for requirements traceability up until 2017.
A subset of this list is relevant as requirement-and-code datasets.
Disregarding publicly unavailable datasets, e.g., non-disclosable industry data sets, and infrequently used datasets, the remaining datasets embody a trend towards standard datasets being re-used as baselines and for better comparability\footnote{We show a tabular view of this list in our supplementary material.}.
These standard datasets including iTrust, eTour, SMOS, and eANCI, indeed are frequently used in more recent studies~\cite{wang2026r2code, jinUserTrace2025, fuchbeta2025lissa, moran2020improving, ali2024establishing, hey2021improving}.
The code-and-requirement datasets provide a golden set of requirements and their trace links to code that were manually crafted and hence, are usually small in size.
Even combined, they do not suffice to provide sufficient amounts of training data for \textit{Deep Learning (DL)} methods~\cite{lin2021traceability, el2025synthline}. Additionally, these ideal requirements-and-code datasets deviate from  the mixed-quality in real-world requirement datasets~\cite{yang2026assessing}.
Hence, we hypothesized that datasets are more realistic if we use synthetically introduce requirements smells.

\subsection{Requirements Generation from Code}

Xu et al.~\cite{xu2024bridging} show the feasibility of generating requirements from source code.
They use existing requirement-to-code datasets to finetune GPT-models and then generate requirements for the application of legacy systems.
Dearstyne~\cite{dearstyne2025intelligent} tested the generation of requirements from source code as one of the four applied problems in requirements traceability.
Practitioners find that automatically generated requirements are comparable in quality to their handcrafted requirements.
Persson~et ~ al.~\cite{perssoncode2req} propose Code2Req, which allows users to generate requirements from code.
Their approach is based on RAG-supported LLMs and includes existing requirements sets as context information.
They also show the feasibility of automatic requirement generation from code using LLMs, although they describe it as not yet practically valuable.
Jin~et~al.~\cite{jinUserTrace2025} focus on generating requirements from code with the aim of facilitating the understanding and verification of code generated by LLMs. 

These approaches do not consider two major aspects:
First, they aim to produce exclusively high-quality requirements, rather than generating a dataset that reflects the characteristics of real-world requirements, which are of mixed quality~\cite{yang2026assessing}.
Dearstyne's work tries to correct for that by incorporating existing sets of requirements from the given codebase as context to the LLMs.
However, this is not enough since (i) an idealized set of context requirements leads to the initial problem of unrealistic requirements, and (ii) a realistic set of context requirements leads to an inexplicit quality decrease in output requirements, as reported in her work.
Without explicitly knowing the location of these qualitative deficiencies, they are of little use to research.
Second, their approaches are fully automated without human validation or feedback. 
In contrast, when producing datasets intended for downstream research or project work, an additional verification and validation step is required to ensure the structural and content correctness of the resulting artifacts.
This human feedback can also be used to guide the requirements generation process.




\section{Approach} \label{sec:methodology}

\begin{figure*}
    \centering
    \includegraphics[width=.85\textwidth]{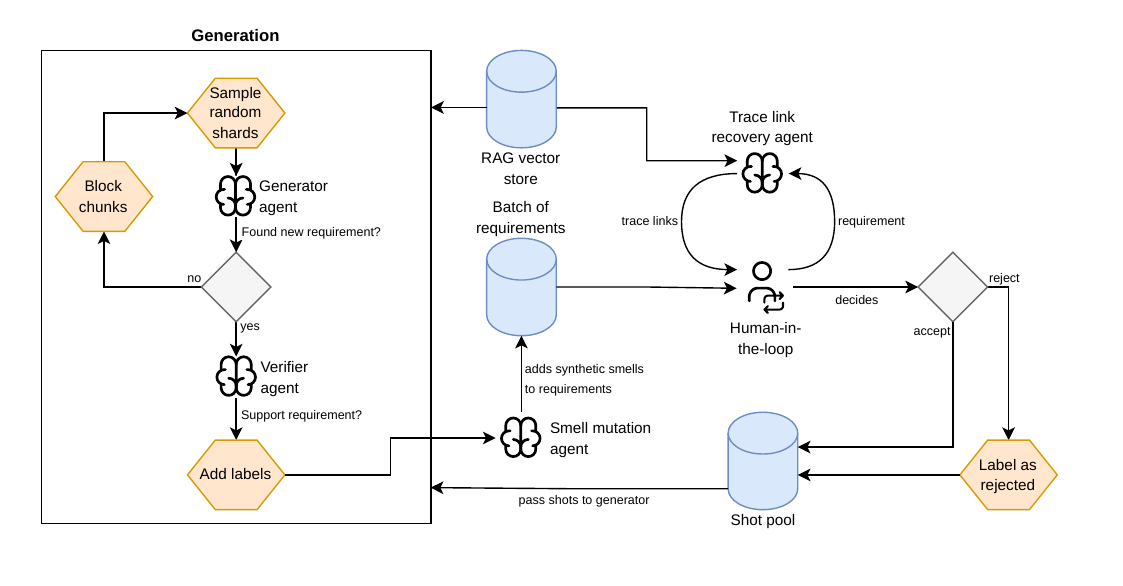}
    \caption{An overview of the complete agentic architecture of our experimental approach.}
    \label{fig:architecture}
\end{figure*}

This section describes the approach used for our study. Figure~\ref{fig:architecture} illustrates the complete architecture.

\subsection{Agentic Approach} \label{sec:agentic-approach}

\textbf{Generator Agent.} The generator agent is responsible for producing batches of $n$ requirements (in our experiments: $n = 10$). To accomplish this, the agent employs RAG to retrieve code chunks from the target project's code base. For each batch, we invoke the generator separately $n$ times, each time producing a \textit{candidate} requirement. The system prompt deliberately refrains from requiring the generated requirement to be guaranteed as implemented or valid, as this is the responsibility of the subsequent verifier agent. Instead, the prompt provides guidance on the characteristics of well-formed requirements and conventions for how requirements should be expressed. Furthermore, the LLM is instructed to output a rationale explaining the basis for the candidate requirement, the code files deemed relevant, the names of the functions that supposedly implement the requirement, and a confidence value. The user prompt supplies the LLM with a list of all previously generated requirements, labeled by the human-in-the-loop (cf. Section~\ref{sec:human-in-the-loop}). The exact prompts are provided in the supplementary material$^1$.

As RAG returns only the most relevant results for a given query, we employ \textit{sharding} (cf. the work of Shao et al.~\cite{shaoScaling2024}), which enables systematic retrieval of different regions of the code base. Each chunk is assigned a random shard from 00--99 as metadata. At each generator invocation, a random subset of three candidate chunks is selected, restricting retrieval to chunks belonging to that subset. If the generator cannot identify novel requirements within those chunks, it is instructed to return no requirement. In this case, the shard subset is resampled and the generator re-invoked. If no more requirements can be extracted from a chunk, it is marked as \textit{saturated} and excluded from future subsets. We sample the shards randomly to allow for a diverse set of requirements to be generated.

As a control group for evaluation, the generator is additionally tasked to generate a configurable number of non-implemented requirements (20~\% in our experiments). In this case, the prompt instructs the LLM to reconstruct requirements for which the agent does not see clear evidence about their implementation.

\textbf{Verifier Agent.} This agent is tasked with determining whether a previously generated requirement candidate is actually implemented in the code base. Unlike the generator, the verifier is permitted to search the entire code base with RAG, as the most relevant retrieval results are expected to concentrate around the candidate requirement. We prompt the LLM to classify the retrieved evidence as either (a) \textit{supporting}, (b) \textit{contradicting}, or (c) \textit{insufficient} for the candidate requirement. To retrieve a diverse and representative set of evidence chunks, the verifier issues 3 targeted RAG queries:
(i) the top-4 chunks retrieved by the candidate requirement text, capturing 
functional intent, (ii) the top-3 chunks retrieved by the function names identified 
by the generator, targeting likely implementation sites, and (iii) the top-3 chunks retrieved  
by the file names identified by the generator. We determined this query constellation to be productive in initial experiments.

Beyond its verdict (supporting, contradicting, or insufficient), the verifier is also instructed to return a rationale, the relevant files and functions, and a confidence value. The system prompt additionally includes explicit evaluation rules to guide the LLM in distinguishing between the verdict categories. In the user prompt, the verifier is provided with the candidate requirement, and the files, functions, and rationale returned by the generator agent.

\textbf{Mutation Agent.} Analogously to the generation of non-implemented requirements, the proportion of smelly requirements in each batch is configurable (20~\% in our experiments). 
Smelly requirements are generated by selecting requirements verified as implemented by the verifier agent and \textit{mutating} them to synthetically introduce target requirements smells. 
The system prompt requests preserving the original requirements as much as possible, restricting changes strictly to those parts necessary to introduce the target smell. 
In the user prompt, we include the requirement to be mutated, the randomly selected target smell, chosen from a list of smell categories (cf. Table~\ref{tab:smell-categories}), and a description of the smell category.

The agent employs RAG to retrieve requirements similar to the candidate from a curated dataset of smelly requirements, pre-filtered to include only examples exhibiting the selected target smell. These examples then serve as few-shot demonstrations to improve mutation quality. The mutated requirement is added to the batch as a smelly variant of an otherwise well-formed requirement.

\textbf{Trace-Link-Recovery (TLR) Agent.} The TLR agent serves the purpose of
supporting the human-in-the-loop in assessing the implementation status of a 
requirement, thus producing verified trace links for all generated 
requirements. The agent receives a candidate requirement together with the file 
and function names identified by the preceding agents. It then issues targeted 
RAG queries to retrieve relevant code chunks: (i) the top-2 chunks retrieved by 
the requirement text, (ii) the top-4 chunks retrieved by function names, and (iii) 
the top-4 chunks retrieved by file names. Again, this constellation was determined
through earlier experiments.

The agent is prompted to identify the exact lines of code that implement the 
given requirement. To support this, the system prompt includes explicit tracing 
guidelines specifying that trace links must be evidence-based. Clear 
criteria are established for what constitutes a valid trace link, including 
conditions, assignments, and method signatures, as well as for what must be 
excluded, such as calls to other methods, comments, and tests. These guidelines 
closely follow the tracing rules by Vogelsang et al.~\cite{vogelsangImpact2025}.

\subsection{Retrieval-Augmented Generation (RAG)} \label{sec:rag}

We employ RAG to enable the agent to efficiently access the complete code base of the target projects. We preprocess the source code of each project prior to indexing. First, we exclude all files irrelevant to the approach, i.e., non-code files and test files. The remaining files are then split into chunks of 100 lines each, with an overlap of 10 lines at the beginning and 10 lines at the end of each chunk. This serves the purpose of balancing retrieval precision against context coverage, as overly large chunks risk introducing excessive irrelevant code into the results.

We prefix each chunk with a metadata header containing the file identifier, the complete path within the project, and the line range covered by the chunk. This facilitates file-based retrieval, as the file path can be retrieved directly from the chunk text. In addition, each chunk is assigned a random shard identifier (cf.~\ref{sec:agentic-approach}), which is added to the vector store as metadata. For the TLR agent, each chunk is additionally stored in a second variant with line numbers prepended to every line of code. This allows the TLR agent to return exact line numbers reliably.

\subsection{Human-in-the-loop \& Multi-shot Prompting} \label{sec:human-in-the-loop}

We follow a human-in-the-loop approach for two reasons. First, because we generate requirements from source code without pre-existing ground truth, automated verification alone does not guarantee validity. Second, Unterbusch and Vogelsang~\cite{unterbuschContextAdaptive2026} demonstrate that validated human reasoning in the RE context, incorporated via few-shot learning, enables LLMs to adapt to context-specific quality judgments efficiently. This suggests that human involvement grounds model behavior in ways that purely automated approaches cannot replicate.

We feed back all requirements labeled by the human-in-the-loop to the generator agent as few-shot examples. This feedback loop serves the purpose of LLM adaptation as described by Unterbusch et al.
Additionally, it also prevents the generation of 
duplicate requirements by making the generator aware of all 
previously accepted (or rejected) requirements.

\section{Study Design}

To report the study design and study results, we follow the guidelines by Jedlitschka and Pfahl~\cite{jedlitschkaReporting2005} on how to report experiments in software engineering.

\subsection{Goals \& Research Questions}

The goal of our human-in-the-loop study is to explore whether the generation of realistic synthetic requirements from source code is viable. 
We assess this in three areas.
First, we investigate whether the requirements generated by our approach are actually implemented in the code base and not hallucinated. Human-verified trace links act as ground truth. 
Second, we evaluate the quality of the generated requirements by examining whether they are free of requirements smells.
Third, we investigate the viability of synthetically introducing requirements smells to produce realistic datasets.
This results in the following research questions:

\begin{compactitem}
    \item \textbf{RQ1.1:} To what extent are requirements reconstructed from source code by an LLM actually implemented in the code base? \textit{(Implementation Accuracy)}
    \item \textbf{RQ1.2:} To what extent are the reconstructed requirements hallucinated by an LLM?
    \textit{(Hallucination Rate)}
    \item \textbf{RQ2:} To what extent are requirements generated by an LLM free of requirements smells?
    \textit{(Generation Quality)}
    \item \textbf{RQ3:} To what extent do human evaluators agree with the LLM on the category of synthetically introduced requirements smells?
    \textit{(Human-LLM Reliability)}
\end{compactitem}

\subsection{Datasets \& Participants}

We use two software projects as code bases to generate synthetic requirements. 
The first project, hereafter referred to as \textit{SEP}, 
is a student project developed by bachelor's students during a practical software engineering course at our university. 
The source code comprises approximately 24,000 lines of code written in Java, TypeScript, HTML, and CSS, and implements a full-stack web application resembling a personal driver hiring platform (like UBER).
The second project, \textit{Mattermost}, is the publicly available source code\footnote{\url{https://github.com/mattermost/desktop} (commit \texttt{adfcd2c} on \texttt{main}).} of the desktop client of the Mattermost messaging application. 
It has approximately 28,000 lines of code written in TypeScript and HTML.
The two projects were selected to cover a diverse range of code characteristics, ranging from a well-maintained long-term open-source project to a student project developed under less rigorous review practices within a constrained timeframe.

All human evaluators involved are PhD students with experience in Requirements Engineering.

\subsection{Experimental Setup} \label{sec:experimental-setup}

\begin{table*}
    \centering
    \caption{Smell Categories Used Throughout the Study. Extracted From Previous Work by Vogelsang et al.~\cite{vogelsangImpact2025} and Extended for This Paper Based on Work by Frattini et al.~\cite{li2024llm}.}
    \begin{tabularx}{\linewidth}{@{}p{2.5cm} X@{}}
        \toprule
        \textbf{Smell Category} & \textbf{Description} \\ \midrule
        Subjective Language & Words of which the semantics are not objectively defined, such as user friendly, easy to use, cost effective, etc. OR Sentences expressing personal opinions or feelings. \\
        Ambiguities & Sentences or sentence parts that are unclear/imprecise and can be misunderstood if read by different people. \\
        Loopholes & Sentences containing escape clauses or conditional exceptions that allow the requirement to be circumvented or interpreted as not applicable, e.g., using phrases like 'where possible', 'as appropriate', 'unless otherwise agreed', etc. \\
        Open-ended,\newline non-verifiable terms & Terms or phrases that cannot be objectively measured or tested, such as 'sufficient', 'adequate', 'reasonable', 'as needed', etc. \\
        Superlatives & Over strong guarantees using absolute words such as always, never, guarantee, 100~\%, cannot fail, etc. \\
        Comparatives & Comparative words such as better, more, etc. \\
        Negative Statements & Sentences containing negative modifiers (e.g., not), negative expressions. \\
        Vague Pronouns & Pronouns that refer back to a previous part of the text for which the reference is unclear. \\
        Non-atomic & Requirements that combine multiple distinct conditions, behaviors, or features into a single statement, making them difficult to test or trace individually. \\
        Passive Voice & Sentences using passive voice such that it is unclear who is performing a certain action. \\
        Optional Parts & Sentences containing optional parts, e.g., by using the words possibly, eventually, if possible, if needed, etc. \\
        Weak Verbs & Weak verbs, such as can, could, may, etc. \\
        UI/UX Aesthetics & Requirements to UI/UX without measurable criteria, such as modern look, pleasant UI, nice animations, etc. \\
        Scope Creep & Making too broad assumptions, such as support all languages, work on all devices, integrate with any service, etc. \\
        \bottomrule
    \end{tabularx}
    \label{tab:smell-categories}
\end{table*}

To conduct our experiments, we implemented the agentic approach described in 
Section~\ref{sec:methodology} in Python. 
To make the approach accessible to evaluators, we developed a web application using \textit{SvelteKit} that interfaces with the Python backend. LLM requests were managed using \textit{Celery} tasks, enabling 
multiple agent invocations to be processed simultaneously. 
During the study, evaluators accessed the locally deployed web application to complete their 
evaluations, with all results stored in a database for subsequent export and analysis. 
For retrieval-augmented generation, we used \textit{ChromaDB} as a local vector store, with embeddings produced by OpenAI's 
\texttt{text-embedding-3-large} model.

We selected OpenAI's \texttt{gpt-5.4-mini-2026-03-17} as the model for all agentic tasks, as it demonstrated suitable performance across all tasks while offering a favorable cost-performance tradeoff compared to the full \texttt{gpt-5.4} model.
We did not evaluate any additional models.

For each of the two projects, three independent evaluators evaluated the generated requirements, serving as the human-in-the-loop for one round of generation. The evaluators were asked to assess requirements for one hour with respect to the following three criteria:
\begin{compactenum}
    \item \textit{Appropriateness.} A subjective 5-point Likert scale score 
    assessing two aspects: (a) whether the requirement conforms to the expected 
    format, and (b) whether it describes plausible system behavior that could 
    be expected from the given system.
    \item \textit{Requirements Smells.} All requirements smells identified by 
    the evaluator in the given requirement, selected from a predefined 
    list of 14 smell types (see Table~\ref{tab:smell-categories}).
    \item \textit{Implementation Status.} Based on the traceability information produced by the TLR 
    agent, highlighting the most relevant code locations related to the 
    requirement's implementation, evaluators classify the requirement as 
    \textit{fully implemented}, \textit{partially implemented}, or \textit{not implemented}.
\end{compactenum}
While the initial design aimed for a single evaluation round (only by the human-in-the-loop), we planned additional rounds as a contingency to resolve potential inter-rater disagreements. In a second round, the evaluators were rotated across 
requirements sets and asked to independently re-rate the appropriateness and smell labels to assess inter-rater agreement. During the rating, they did not have access to the first evaluator's decisions.
In a third round, two authors of this paper reviewed all 
inter-rater disagreements in the smell labels and resolved them by developing a codebook addressing common sources of confusion that arose during evaluation. Although the specific conventions were derived from disagreements observed in practice rather than specified in advance, the following codebook was established to ensure consistent labeling across evaluators.
\begin{compactitem}
    \item The use of \textit{should} alone does not constitute the 
    \textit{weak verb} smell;
    \item the \textit{vague pronoun} smell applies only when multiple 
    possible references exist for a given pronoun;
    \item when both \textit{UI/UX aesthetics} and \textit{subjective language} 
    are applicable, the more specific \textit{UI/UX aesthetics} smell 
    should be used;
    \item the smells \textit{open-ended, non-verifiable terms} and 
    \textit{subjective language} are consolidated into \textit{subjective 
    language};
    \item the \textit{passive voice} smell applies only when the use of 
    passive voice results in an unclear or unidentifiable actor.
\end{compactitem}

\section{Study Results}

\subsection{Human-in-the-loop Datasets}

The human-in-the-loop study produced a total of 188 requirements across the two source code projects. The generated requirements are labeled either as (i) implemented, i.e., implemented and smell-free by the LLM, (ii) non-implemented, i.e., not-implemented and smell-free, or (iii) smelly, i.e., including a synthetically added requirement smell. We introduce the non-implemented class as a control group. These categories are solely based on the LLM's perspective.
We show the distribution of implemented, non-implemented, and smelly requirements in Table~\ref{tab:datasets}.

\begin{table}
    \centering
    \caption{Requirements Generated During the Human-in-the-loop Study}
    \begin{tabularx}{\linewidth}{@{}X c r r r r@{}}
        \toprule
        \multirow{2}{*}{\textbf{Project}} & \multirow{2}{*}{\textbf{Evaluator}} & \multicolumn{4}{c}{\textbf{\# of Requirements}} \\
        & & Impl. & Non-impl. & Smelly & \textbf{Total} \\ \midrule
        \multirow{3}{*}{SEP} & E1 & 18 & 7 & 5 & \textbf{30} \\
        & E2 & 19 & 8 & 7 & \textbf{34} \\
        & E3 & 21 & 7 & 6 & \textbf{34} \\ \midrule
        \multirow{3}{*}{Mattermost} & E1 & 16 & 5 & 5 & \textbf{26} \\
        & E2 & 21 & 6 & 5 & \textbf{32} \\
        & E3 & 19 & 8 & 5 & \textbf{32} \\
        \bottomrule
    \end{tabularx}
    \label{tab:datasets}
\end{table}

\subsection{RQ1.1: Implementation Accuracy}

This research question aims to assess the implementation accuracy of the 
LLM-generated requirements by examining the extent to which requirements 
generated from source code are confirmed as implemented by the human-in-the-loop evaluators, 
based on the trace link code excerpts presented to them during the study. We report precision, recall, and $F_1$-scores of comparing the LLM's generation methods (i.e., whether it was tasked to generate an implemented or non-implemented requirement) with the labels of the human-in-the-loop evaluators in Table~\ref{tab:rq1-1-results}. To compute these values, we define true positives as labeled implemented by both the LLM and human evaluator, false positives as labeled implemented by the LLM but labeled non-implemented by the human, etc.

As can be seen, there is a consistent gap between precision and recall across all evaluators and projects. While the precision is consistently high (0.857--1.000), the recall stays lower (0.708--0.769). When the LLM is tasked to generate an implemented requirement, humans almost always agree.
Still, when the LLM is tasked to generate a non-implemented requirement deliberately, humans often find it to be implemented based on the given trace links. 
Furthermore, the performance for Mattermost is consistently stronger than for SEP. A possible explanation for this may be the higher code quality of a well-maintained long-term project.

\begin{table}
    \centering
    \caption{RQ1.1: Implementation Accuracy of the LLM Based on Human-In-The-Loop Labeling}
    \begin{tabularx}{\linewidth}{@{}X c r r r@{}}
        \toprule
        \multirow{2}{*}{\textbf{Project}} & \multirow{2}{*}{\textbf{Evaluator}} & \multicolumn{3}{c}{\textbf{Results}} \\
        & & Precision & Recall & $F_1$-score \\ \midrule
        \multirow{4}{*}{SEP} & E1 & 0.944 & 0.708 & 0.810 \\
        & E2 & 0.895 & 0.739 & 0.810 \\
        & E3 & 0.857 & 0.720 & 0.783 \\
        & \textbf{\textit{E1--E3}} & \textbf{\textit{0.897}} & \textbf{\textit{0.722}} & \textbf{\textit{0.800}} \\ \midrule
        \multirow{4}{*}{Mattermost} & E1 & 1.000 & 0.762 & 0.865 \\
        & E2 & 0.952 & 0.769 & 0.851 \\
        & E3 & 1.000 & 0.731 & 0.844 \\
        & \textbf{\textit{E1--E3}} & \textbf{\textit{0.982}} & \textbf{\textit{0.753}} & \textbf{\textit{0.853}} \\ \midrule
        \textbf{All} & & \textbf{0.939} & \textbf{0.738} & \textbf{0.826} \\
        \bottomrule
    \end{tabularx}
    \label{tab:rq1-1-results}
\end{table}

\subsection{RQ1.2: Hallucination Rate}

In Table~\ref{tab:rq1-2-results}, we report the false positive requirements (i.e., labeled as implemented by the LLM and labeled as non-implemented by the human-in-the-loop), as well as the \textit{hallucination rate}, which is defined as the false positive rate.

The overall hallucination rate of 6.1~\% is low, which suggests that the LLM reliably generates requirements that are actually implemented. Still, there is a large difference between the two projects. For Mattermost, we see a rate of only 1.8~\%, while for SEP, the rate is 10.3~\%. This discrepancy follows the observation from RQ1.1.
It is possible, that the code quality is relevant for this difference. 
Well-structured code that has been developed over a long time-span with many different reviewers, eases the LLM to ground the generated requirements in clear evidence.
However, as Mattermost is also more complex than the student projects, it might have been easier for the evaluators to identify hallucinations as such.
Though, as there are only 7 reported hallucinations in total, generalizability is limited.


\begin{table}
    \centering
    \caption{RQ1.2: Requirements Hallucinated by the LLM and Hallucination Rate as Assessed by the Human-In-The-Loop Using Trace Links}
    \begin{tabularx}{\linewidth}{@{}X r r r@{}}
        \toprule
        \multirow{2}{*}{\textbf{Project}} & \multicolumn{3}{c}{\textbf{Results}} \\
        & LLM: Impl. & FP (Hallucinated) & Hallucination Rate \\ \midrule
        SEP & 58 & 6 & 10.3~\% \\
        Mattermost & 56 & 1 & 1.8~\% \\
        \textbf{Total} & \textbf{114} & \textbf{7} & \textbf{6.1~\%} \\
        \bottomrule
    \end{tabularx}
    \label{tab:rq1-2-results}
\end{table}

\subsection{RQ2: Generation Quality}

Comparing the smell labels of the first-round human-in-the-loop with the second-round evaluators, we find that Cohen's $\kappa$ ranges from $-0.083$ (Mattermost; E1) to $0.803$ (Mattermost; E3), with an overall $\kappa = 0.213$ across both projects and all evaluator pairs, indicating fair agreement at best. 
This disagreement shows the necessity of the third labeling round, in which two authors of this paper resolved disagreements by developing a codebook from common sources of confusion (cf. Section~\ref{sec:experimental-setup}). The peer-reviewed labels from this third round are used as the basis for the results reported below. We report all results in Table~\ref{tab:rq2-results}.

The smell rate in requirements labeled as smell-free (implemented and non-implemented combined) by the LLM, over both projects and all evaluators, is 24.5~\%.
This smell rate is calculated based on the peer-reviewed smells. As observed for RQ1.1 and RQ1.2, both projects exhibit notable differences in variation. The smell rates for SEP range from 7.1~\%--40.7~\%, showing considerable variance across the 3 evaluators. 
For Mattermost, the results are much more consistent with a range of 25.9~\%--33.3~\%. 

Of all possible smell categories, only 6 categories were labeled to appear in the allegedly smell-free requirements. These categories, in descending order of frequency, are: \textit{non-atomic} (12), \textit{passive voice} (11), \textit{ambiguities} (8), \textit{negative statements} (6), \textit{subjective language} (3), and \textit{vague pronouns} (1).
Note, that a single requirement may contain multiple smells. Many of these dominant smells are structural and syntactic, rather than semantic ones, suggesting that the LLM tends to generate requirements that are functionally correct but linguistically imprecise. 
The absence of smell categories such as \textit{superlatives} or \textit{comparatives} together with the functional correctness suggests that the LLM does well in presenting the code behavior based on the given information.

A representative example for a non-atomic requirement as generated by the LLM is (Mattermost): \textit{''When a download item no longer has a valid file location, opening its folder should mark it as deleted and, if a download location is configured, open that download folder instead.''} This highlights the LLM trying to include closely related functionality (in code) in a single requirement, leading to long, complicated, and/or non-atomic requirements. 
Another representative requirement, this time for passive voice, is: \textit{''When the main window is closed on Windows or Linux and tray minimization is enabled, the application hides the window instead of quitting.''} 
It is unclear which actor is responsible for closing the main window.
Ambiguous requirements would often allow various interpretations of the intended functionality, e.g., in: \textit{''If a notification sound is configured, the notification is shown silently and the chosen sound is stored for later playback''} it is unclear whether the mentioned silent notification is shown directly after configuring a notification sound, or whether this configures a general system property.

\begin{table}
    \centering
    \caption{RQ2: Requirements Smells Found in LLM-Generated Smell-Free Requirements After Peer-Review}
    \begin{tabularx}{\linewidth}{@{}X c r r r@{}}
        \toprule
        \multirow{2}{*}{\textbf{Project}} & \multirow{2}{*}{\textbf{Evaluator}} & \multicolumn{3}{c}{\textbf{Results}} \\
        & & \# of Reqs. & \# of Smelly Reqs. & Smell Rate \\ \midrule
        \multirow{4}{*}{SEP} & E1 & 25 & 3 & 12.0~\% \\
        & E2 & 27 & 11 & 40.7~\% \\
        & E3 & 28 & 2 & 7.1~\% \\
        & \textbf{\textit{E1--E3}} & \textbf{\textit{80}} & \textbf{\textit{16}} & \textbf{\textit{20.0~\%}} \\ \midrule
        \multirow{4}{*}{Mattermost} & E1 & 21 & 7 & 33.3~\% \\
        & E2 & 27 & 8 & 29.6~\% \\
        & E3 & 27 & 7 & 25.9~\% \\
        & \textbf{\textit{E1--E3}} & \textbf{\textit{75}} & \textbf{\textit{22}} & \textbf{\textit{29.3~\%}} \\ \midrule
        \textbf{All} & & \textbf{155} & \textbf{38} & \textbf{24.5~\%} \\
        \bottomrule
    \end{tabularx}
    \label{tab:rq2-results}
\end{table}

\subsection{RQ3: Human-LLM Reliability}

To assess whether synthetically introduced smells are reliably detected by human evaluators, we compare the LLM's intended smells against the peer-reviewed smells from the final evaluation round. In both projects, 60.6~\% of intentionally smelly requirements were labeled with the exact intended smell category, and SEP showed a higher exact-match agreement (66.7~\%) than Mattermost (53.3~\%). When considering whether any smell was detected, regardless of category, 90.9~\% of intentionally smelly requirements received at least one smell label. 
We report all results in Table~\ref{tab:rq3-results}.

\begin{figure}
    \centering
    \includegraphics[width=0.9\linewidth]{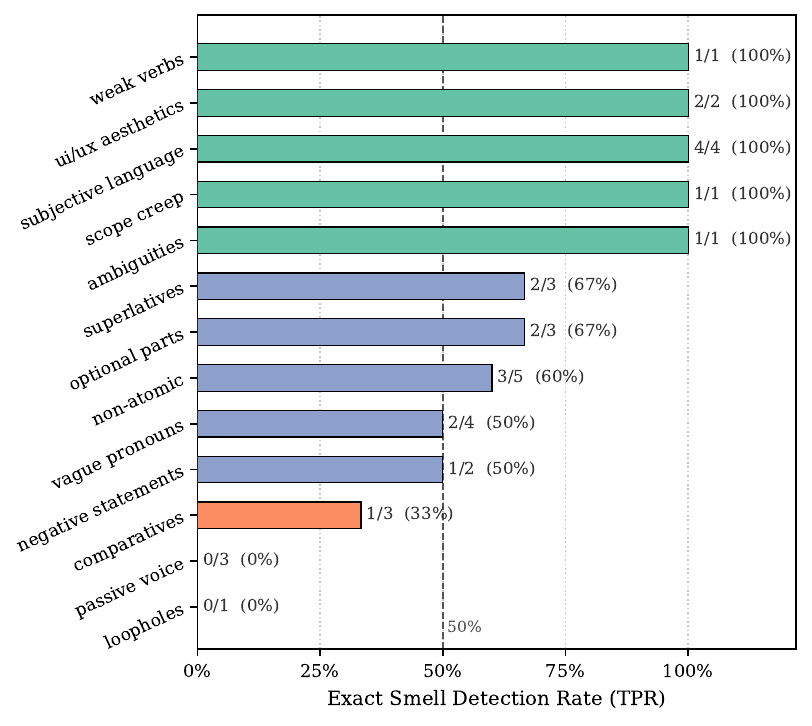}
    \caption{Agreement of human evaluators (after peer-review) with LLM on the categories of synthetically introduced smells.}
    \label{fig:rq3-2-results}
\end{figure}

Figure~\ref{fig:rq3-2-results} shows the per-category agreement: while some smells, e.g., \textit{subjective language} and \textit{UI/UX aesthetics}, are detected with 
perfect agreement, other smell categories such as \textit{passive 
voice} and \textit{loopholes} show zero exact-match agreement.

\begin{table}
    \centering
    \caption{RQ3: Requirements Smells Found in LLM-Generated Smelly Requirements After Peer-Review}
    \begin{tabularx}{\linewidth}{@{}X c r r r r r@{}}
        \toprule
        \multirow{2}{*}{\textbf{Project}} & \multirow{2}{*}{\textbf{Evaluator}} & \multirow{2}{1.5cm}{\# of Smelly Reqs.} & \multicolumn{2}{c}{\textbf{Any Smell?}} & \multicolumn{2}{c}{\textbf{Exact Smells?}} \\
        & & & \# & \% & \# & \% \\ \midrule
        \multirow{4}{*}{SEP} & E1 & 5 & 4 & 80.0~\% & 4 & 80.0~\% \\
        & E2 & 7 & 7 & 100.0~\% & 5 & 71.4~\% \\
        & E3 & 6 & 5 & 83.3~\% & 3 & 50.0~\% \\
        & \textbf{\textit{E1--E3}} & \textbf{\textit{18}} & \textbf{\textit{16}} & \textbf{\textit{88.9~\%}} & \textbf{\textit{12}} & \textbf{\textit{66.7~\%}} \\ \midrule
        \multirow{4}{*}{Mattermost} & E1 & 5 & 5 & 100.0~\% & 2 & 40.0~\% \\
        & E2 & 5 & 5 & 100.0~\% & 3 & 60.0~\% \\
        & E3 & 5 & 4 & 80.0~\% & 3 & 60.0~\% \\
        & \textbf{\textit{E1--E3}} & \textbf{\textit{15}} & \textbf{\textit{14}} & \textbf{\textit{93.3~\%}} & \textbf{\textit{8}} & \textbf{\textit{53.3~\%}} \\ \midrule
        \textbf{All} & & \textbf{33} & \textbf{30} & \textbf{90.9~\%} & \textbf{20} & \textbf{60.6~\%} \\
        \bottomrule
    \end{tabularx}
    \label{tab:rq3-results}
\end{table}

\subsection{Appropriateness}

Lastly, we evaluated the appropriateness scores. We did not include them as a 
research question, as they are a highly subjective measure, which is reflected in the inter-rater agreement between the first and second evaluation rounds. 
Over both projects, 
Krippendorff's $\alpha = 0.178$, with an exact agreement of 36.2~\% (70.2~\% 
when allowing a one-point deviation between ratings), confirms that 
appropriateness scores are inconsistent across evaluators and should be 
interpreted accordingly.

Comparing appropriateness scores across generation methods, the Kruskal-Wallis 
test reveals a significant overall effect for Mattermost ($p = 0.0001$; $\varepsilon^2 = 0.212$) and across both projects combined ($p = 0.0034$; $\varepsilon^2 = 0.061$), 
but not for SEP alone ($p = 0.0981$; $\varepsilon^2 = 0.048$). Pairwise comparisons using the Mann-Whitney U test with Holm--Bonferroni correction show 
that for Mattermost, positive requirements score significantly higher than both 
not-implemented ($p_{\text{adj}} = 0.0043$, $p < 0.01$; $r = -0.447$) 
and smelly requirements ($p_{\text{adj}} = 0.0005$, 
$p < 0.001$; $r = -0.598$). For SEP, we do not detect significant differences between generation methods after Holm--Bonferroni correction.
Across both projects combined, only the contrast between positive and smelly 
requirements is calculated as significant ($p_{\text{adj}} = 0.002$, $p < 0.01$; $r = -0.367$), while non-implemented requirements do not differ significantly from either group. These results suggest that human 
evaluators likely perceive smelly requirements as less appropriate than well-formed ones.

\begin{figure*}
    \centering
    \includegraphics[width=0.8\textwidth]{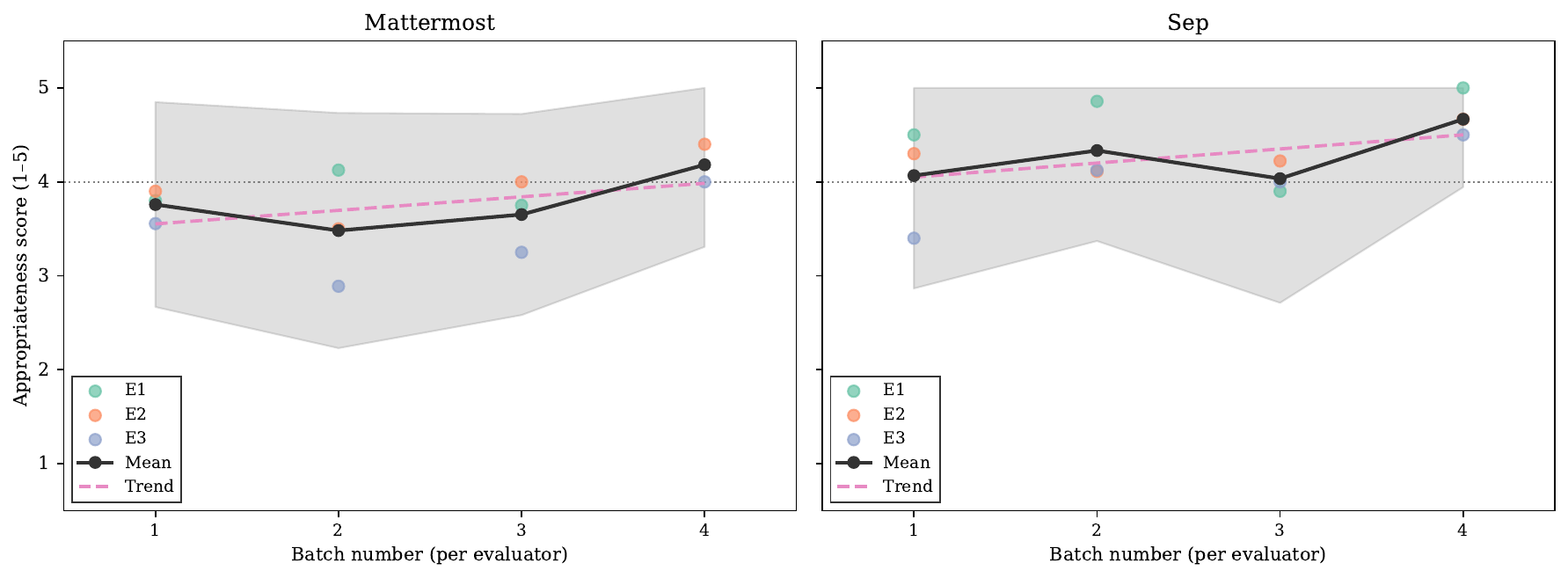}
    \caption{Comparison of human-in-the-loop appropriateness ratings over time. Each batch is generated with reviews from last batch as shot samples.}
    \label{fig:appropriateness-over-batches}
\end{figure*}

Figure~\ref{fig:appropriateness-over-batches} shows the mean appropriateness 
scores per batch between evaluators for both projects. Although the trend line 
suggests a slight upward trajectory over successive batches, indicating that the multi-shot feedback loop may have an impact, for both projects, this correlation is not statistically significant (Spearman's rank correlation: $\rho = 0.40$; $p = 0.600$), 
and the small number of batches ($n = 4$) limits any meaningful interpretation 
of this trend. 
We note it as a tentative observation that warrants further investigation with larger batch counts.

\section{Discussion}

We initially intended to provide a complete approach for generating 
realistic requirements-to-code datasets.
Throughout the study and evaluation, various problems arose, leading to this 
experience report.
While the hallucination rate is low, confirming our initial hypothesis that, through code-grounding, LLMs should be able to generate implemented requirements, this does not hold true for non-implemented ones.
Low recall comparing the human judgments to the LLM's task shows that, even when specifically prompted to generate non-implemented requirements, the LLM cannot reliably perform this task.
We agree with previous work~\cite{jinUserTrace2025, perssoncode2req} that LLMs show promising results in requirements generation from code. However, similarly to what these studies already hint at, we find that human oversight is still necessary. 

We hypothesized that through clear guidelines provided in the prompt, the LLM 
would be able to generate high-quality requirements if desired, thus containing 
a low rate of unintentional requirements smells.
Seeing that 24.5~\% of requirements are still labeled as smelly by the human 
evaluators also proves this hypothesis wrong.
Femmer and Vogelsang~\cite{femmerRequirements2019} argue that requirements quality is fundamentally quality-in-use: a requirement is only defective if it actually impairs a downstream activity, and whether a given smell does so 
depends heavily on context.
Frattini~\cite{frattiniIdentifying2024} reinforces this with industrial 
evidence, showing that the factors that determine whether a smell constitutes an 
actual defect are numerous and context-specific.
Consequently, the 24.5~\% smell rate reported in RQ2 should be interpreted as 
the rate at which generated requirements exhibit linguistic patterns associated 
with \textit{potential} quality issues, rather than confirmed defects.
Preliminary evidence that smells affect at least some downstream tasks is 
provided by Vogelsang~et~al.~\cite{vogelsangImpact2025}, who show that 
requirements smells negatively impact automated traceability, suggesting that 
this concern is not merely theoretical.

When designing the original study, we assumed that one human-in-the-loop would 
be sufficient to label requirements smells with adequate reliability.
However, the weak inter-rater agreement of $\kappa = 0.213$ made an additional 
round of evaluation and the development of a codebook necessary, defining 
conventions to operationalize context-specific interpretations of smells.
This is supported by the findings of Unterbusch and 
Vogelsang~\cite{unterbuschContextAdaptive2026}, showing that smell judgments 
are context-dependent and not a task that can be objectively completed even by 
trained evaluators.
We acknowledge that this also possibly affects the actual smell rate for RQ2, 
regardless of the countermeasures taken (cf.\ Section~\ref{sec:threats}).

Furthermore, we assumed that synthetically introduced smells could be reliably labeled, enabling the production of realistic datasets.
We see that human evaluators identified a smelly requirements in 90.9~\% of cases, showing that these requirements are recognized as problematic.
However, since only 60.6~\% of smells were exactly identified, the exact smell 
perceived by humans again appears to be context-dependent and not objectively 
identifiable.

For practitioners, our findings suggest that while a low hallucination rate 
indicates that the approach is useful for generating requirements from existing 
codebases, there are other issues that are not immediately obvious.
Generated requirements should not be used without human review, as the LLM 
fails to self-assess quality and to follow the prompted tasks precisely.
Critically, the performance gap between Mattermost and SEP indicates that the 
approach performs better for well-maintained codebases, precisely those 
most in need of requirements generation in practice.
Further investigation of the unintentional smells introduced by LLMs could 
inform prompt engineering improvements, highlighting the defects to which LLMs are 
most vulnerable.

\subsection{Lessons Learned}

From the study and its findings, we learn the following lessons.
\begin{compactenum}
    \item Labeling smells in requirements always requires multiple humans, as smells are highly context-sensitive. Neither an LLM nor a single human evaluator alone is reliable for this task.
    \item LLMs can support the generation of requirements from code, but should not be trusted with this task without human supervision as (a) prompts are not always followed precisely (e.g., generating implemented requirements even when explicitly prompted not to do so) and (b) generated requirements can exhibit major quality deficiencies.
    \item The reliability of LLM-based requirement generation, even when based on source code, strongly depends on the codebase; practitioners should not expect consistent results across codebases of different maintenance levels.
\end{compactenum}

\subsection{Threats to Validity}
\label{sec:threats}

\textbf{Internal Validity.} Prompting must be considered a threat to validity whenever LLMs are employed (cf.\ Korn et al.~\cite{kornReporting2026}), as prompts may miss important details or bias the LLM toward outcomes anticipated by the prompt author. 
We mitigated this by iteratively refining the prompts throughout the study. 
The multi-shot feedback loop introduces a further threat, as each evaluator's later ratings are shaped by their own prior decisions, causing individual biases to accumulate, thus complicating cross-evaluator comparison.
Additionally, since smelly requirements are mutations of previously accepted requirements, evaluators may recognize them based on familiarity with the unmodified counterpart. 
Presenting smell categories as a predefined list may further encourage evaluators to label smells they would not otherwise have noticed, potentially inflating false positive rates.
The evaluator pool consisted exclusively of PhD students with varying levels of RE expertise, which may introduce selection bias and noise, particularly for smell detection.

\textbf{External Validity.} The approach is evaluated exclusively on general-purpose software projects, excluding domains with domain-specific requirements such as safety-critical or heavily regulated systems.
We consider this a limited threat, as the approach relies solely on information retrieved via RAG from the target code base without incorporating domain-specific knowledge. 
Furthermore, the fixed RAG configuration may require reconfiguration for projects that differ substantially in scale from those used in this study.
However, as we only used 2 very different software projects, generalizability of the assumed explanations between Mattermost and SEP is very limited.

\textbf{Construct Validity.} Appropriateness ratings are inherently subjective as evaluators may interpret the 5-point scale differently, leading to inconsistent ratings. 
The smell labels in round three are not purely independent human judgments, as the authors had to resolve disagreements and produce the codebook; the labels therefore may partially reflect the authors' interpretation. 
Implementation status is assessed solely based on TLR agent trace links, meaning that retrieval failures may produce mislabeled requirements rather than reflecting true implementation status. 
We consider this an acceptable tradeoff. 
Finally, results are based on a single LLM; alternative models may yield different outcomes, although the exact model is reported in full to enable reproduction and comparison.

\textbf{Conclusion Validity.} The datasets produced during the study are relatively small, which can reduce statistical power for subgroup analysis.
Although the Holm-Bonferroni correction is applied to each family of comparisons, the overall family-wise error rate across all analyzes may not be fully controllable.

\section{Conclusion}

We present an LLM-based and RAG-supported agentic approach for generating realistic requirements-to-code datasets from existing code bases, and evaluate it in a human-in-the-loop study across two software projects.
Although we originally intended to offer this approach as a viable solution for generating datasets for downstream RE research, we now offer an experience report, as during the study various problems arose.
Although we initially hypothesized that LLMs would be able to reliably generate both implemented and intentionally non-implemented requirements, only the former held true (6.1~\% hallucination rate).
Furthermore, we hypothesized that through prompting with rigorous guidelines, the LLM would be able to generate requirements free of requirements smells (thus potentially high in quality).
In practice, 24.5~\% of the generated requirements exhibit requirements smells, showing that quality cannot be self-assessed by the LLM alone.
A single human-in-the-loop also proved to be insufficient in detecting smells reliably, confirming that requirements smells are context-dependent and cannot be classified reliably without shared conventions, as reflected in a low initial inter-rater agreement of $\kappa = 0.213$.
This required multiple rounds of evaluations that were aimed at resolving common misunderstandings.
Finally, the reliability of LLM-based requirements generation strongly depends on code quality, such that practitioners should not expect consistent results across code bases of different maintenance levels.

These findings suggest that fully autonomous generation of requirements from code is not yet viable, requiring human supervisor at least for quality assurance.
We offer our lessons learned from this study and publish two generated requirements-to-code datasets as a concrete contribution to the empirical RE community.

\subsection{Future Work}

Several directions remain for future work.
Automated smell detection could be applied as an initial optimization step before human 
review, reducing cognitive workload.
Rather than asking evaluators to identify all smells from scratch, the human-in-the-loop could confirm or reject automatically detected smell candidates, shifting the task from open-ended labeling to verification.
The multi-shot feedback loop shows a tentative upward trend in appropriateness scores across batches; a larger evaluation with more batches and evaluators would allow this effect to be examined more rigorously, and may inform whether the feedback loop can reduce the need for human oversight.
Finally, a larger evaluation across more code bases and domains, particularly safety-critical or heavily regulated systems, is necessary to establish the generalizability of these findings.

\section*{Acknowledgments}

We thank the study participants for their time and support throughout 
the evaluation of the presented approach. Large language models were used to 
assist in improving the writing of this paper and to support the development of the 
experimental implementation of the approach.

Funded by the Deutsche Forschungsgemeinschaft (DFG, German Research Foundation) – Project number: 566352773.

\bibliographystyle{IEEEtran}
\bibliography{references}

\end{document}